\documentclass[12pt]{article}

\usepackage{amsmath,amsfonts,latexsym,amssymb,amsthm,graphicx}
\usepackage[margin=1in]{geometry}

\usepackage[round]{natbib}
\usepackage[colorlinks,bookmarksopen,bookmarksnumbered,citecolor=red,
urlcolor=red]{hyperref}
\usepackage{tabularx,ragged2e,booktabs,caption}
\usepackage{subfig}
\graphicspath{{images/}}
\usepackage{setspace}
\usepackage{mathrsfs} 
\usepackage{authblk}

\newcommand{\bfm}[1]{\ensuremath{\mathbf{#1}}}
          \def\cA{{\cal  A}}

     \def\bQ_\lambda{\bfm Q_\lambda}     \def\cQ_\lambda{{\cal  Q_\lambda}}

\renewcommand{\hat}{\widehat}

\def\var{ {\text{var}} }

\def\calA{ {\mathcal A} }

\usepackage{tikz}
\usetikzlibrary{decorations.markings}
\tikzstyle{vertex}=[circle, draw, inner sep=2pt, minimum size=6pt]
\usepackage{pgflibraryarrows} 
\usepackage{pgflibrarysnakes}

\newtheorem{theorem}{Theorem}[section]

 \author[1]{Claudia Di Caterina\thanks{Corresponding author: Piazza Universit\`a 1,
39100 Bolzano, Italy. E-mail: claudia.dicaterina@unibz.it}}
\author[1]{Davide Ferrari}
\affil[1]{Faculty of Economics and Management,  University of Bolzano}
\date{}
 
\title{Sparse composite likelihood selection}

\begin{document}

\maketitle

\begin{abstract}
Composite likelihood has shown promise in settings where the number of parameters $p$ is large due to its ability to break down complex models into simpler components, thus enabling inference even when the full likelihood is not tractable.  Although there are a number of ways to formulate a valid composite likelihood in the finite-$p$ setting, there does not seem to exist agreement on how to construct composite likelihoods that are comp utationally efficient and statistically sound when $p$ is allowed to diverge. This article introduces a method to select sparse composite likelihoods  by minimizing a criterion representing the statistical efficiency of the implied estimator plus an $L_1$-penalty discouraging the inclusion of too many sub-likelihood terms. Conditions under which consistent model selection occurs are studied. Examples illustrating the procedure are analysed in detail and applied to real data.
\end{abstract}

 
\section{Introduction}

While the likelihood function plays a central role in statistics, the widespread availability of large data sets in many fields poses nontrivial challenges to traditional likelihood  methods. Issues related to either specification or computation of the full likelihood  make it difficult to  select  interpretable models and obtain accurate estimates within a reasonable time. These challenges have motivated the development of composite likelihood methods, which avoid the specification of the full likelihood by combining a number of low-dimensional likelihood objects \citep{besag75, Lindsay88}.

Let $Y$ be a $d\times 1$ random vector with density $f(y;\theta)$ indexed by the parameter $\theta\in \Theta \subseteq \mathbb{R}^p$. Suppose that the full $d$-dimensional density of $Y$ is difficult to specify or compute but we can identify $p$  densities  $f_j(y; \theta)$ ($j=1,\dots, p$) defined on low-dimensional subsets of $Y$, such as marginals $Y_j$,  pairs $(Y_j, Y_k)$, or conditionals $Y_j|Y_k=y_k$ ($j\neq  k$). Given independent observations $Y^{(1)}, \dots, Y^{(n)}$ on $Y$, the composite likelihood estimator maximizes the composite log-likelihood function
\begin{align*}
%\label{eq:comp_loglik}
\ell(\theta; Y^{(1)}, \dots, Y^{(n)} ) = \sum_{j=1}^p \ell_j(\theta; Y^{(1)}, \dots, Y^{(n)} )\,, 
\end{align*}
where $\ell_j(\theta; Y^{(1)}, \dots, Y^{(n)} ) = \sum_{i=1}^n \log f_{j}(Y^{(i)}; \theta)$ 
denotes the  sub-likelihood associated with the $j$th data subset.  
The composite likelihood estimator has become  popular  in many areas of statistics due to the simplicity in defining the objective function and computational advantages compared to the maximum likelihood estimator. At the same time, it has the same desirable first-order properties as maximum likelihood, such as consistency; see 
\cite{varin11} for a comprehensive survey.    

The composite likelihood framework naturally suits problems where the parameter dimension $p$ is allowed to diverge with the sample size. Nonetheless, the issue of composite likelihood selection, that is how to select the sub-likelihoods that form the overall composite likelihood \citep{lindsay2011issues}, remains largely unresolved in such a setting. Likelihood selection is crucial since it determines both statistical properties and computing cost of the resulting estimator \citep{cox2004note, xu2011robustness, lindsay2011issues, huang2020specification}; it is also related to  model selection, with the two tasks coinciding when each sub-likelihood contains distinct elements of $\theta$.   Without some form of  selection, the accuracy of common composite likelihood estimators, such as the pairwise likelihood estimator, is shown to deteriorate as the data dimension grows when the low-dimensional data subsets are sufficiently correlated \citep{cox2004note,ferrari2016parsimonious}.

In the finite-$p$ setting,  different selection strategies to balance the trade-off between statistical efficiency and computing cost have been proposed. Typically, instead of using all of the sub-likelihoods, a smaller subset is  selected, albeit determining a good subset remains challenging. \cite{dillon2010stochastic} and \cite{ferrari2016parsimonious} develop stochastic approaches where sub-likelihoods are sampled according to a statistical efficiency criterion. For data with a spatial or temporal structure, sub-likelihoods corresponding to nearby pairs of observations are often considered for practical purposes; e.g., see \cite{heagerty1998composite, sang2014tapered}. For the large-$p$ setting with sparse $\theta$, there does not seem to exist a universal rule for sub-likelihood selection that is statistically justified and computationally convenient. 

This paper introduces a flexible and computationally convenient method to build a composite likelihood function starting from a very large number of potential sub-likelihood candidates.  The main idea is to minimize a  convex criterion representing  statistical efficiency with the addition of a weighted $L_1$-penalty to avoid selection  of too many noisy terms. Each sub-likelihood is assumed to contain distinct elements of $\theta$ in our setting; while this simplification has further computational advantages when $p$ is  large, it also enables one to conduct model selection.  Building on the recent success of shrinkage methods for the full likelihood, many  works have extended the use of sparsity-inducing penalties in the composite likelihood framework for particular models; e.g., see \cite{bradic2011penalized, xue2012nonconcave,  gao2017data}. The  approach considered here is entirely different from these methods, since our penalty focuses on selection of sub-likelihood functions rather than of elements of $\theta$. Differently from classic shrinkage methods, our strategy has the advantage to retain unbiasedness of the final estimating equations and consistency of the related parameter estimator.

\section{Methods} \label{sec:methods}

\subsection{Sparse composite likelihood selection}

Let us focus on the case where the parameter vector $\theta = (\theta_1, \dots, \theta_p)^\top$ is sparse, in the sense that a large fraction of its elements are exactly zero, and $p$ is allowed to grow  with the sample size $n$.  Throughout the paper,  $\cA =\{ j: \theta_{j} \neq 0 \}$ is   the index set for the $p^\ast$ nonzero elements in $\theta$ and indicates respective sub-vectors and sub-matrices when used as a subscript. It is assumed here that each  sub-likelihood  $\ell_j(\theta)$   depends only on the specific component $\theta_j $; this simplification has computational advantages when $p$ is large. The marginal scores are defined by $u_j(\theta_j; y) = \partial \log f_j(y; \theta_j)/ \partial \theta_j$ ($j=1, \dots, p$), whilst $u(\theta; y) = \{u_1(\theta_1; y) , \dots, u_p(\theta_p; y)\}^\top$  denotes the vector collecting all these scores. The approach presented next is also valid for the more general setting  where each sub-likelihood depends on a finite number of parameters, in which case the $j$th score equals $u_{j}(\theta; y) = \sum_{k=1}^p \partial \log f_k(y; \theta)/ \partial \theta_j$.

The main goal is to reduce the model dimension by dropping all the zero elements of $\theta$ while estimating the remaining elements. To this end, we take the estimator $\hat \theta$ with $j$th element defined by $\hat \theta_j = \tilde \theta_j I(\hat w_j \neq 0)$,
where $\tilde \theta_j$ is the $j$th marginal estimator
\begin{equation*}  %\label{eq:thetatilde}
\tilde \theta_{j} =  \left\{ 
\theta_j : 0=   \sum_{i=1}^n u_{j}(\theta_{j}; Y^{(i)}) \right\}   \quad (j=1,\dots,p)\,,
\end{equation*}
and  $\hat w = (\hat w_{1}, \dots, \hat w_{p} )^\top$ is the selection rule obtained by minimizing the penalized   objective
\begin{align} \label{eq:criterion_empirical}
\hat d_\lambda(w)  & = \dfrac{1}{2}w^\top \hat C     
w -  w^\top \text{diag}(  \hat C  ) +    \dfrac{\lambda}{n}  \sum_{j=1}^p  \dfrac{     |w_{j}|}{\tilde \theta^2_j},
\end{align}
for some user-specified constant $\lambda \geq 0$. Here $\text{diag}(A)$ is the diagonal vector of the square matrix $A$,   where $\hat C$ is an estimator of the $p\times p$ score covariance matrix $C(\theta) = \var\{u(\theta; Y)\} = E\{u(\theta; Y) u(\theta; Y)^\top \}$. A natural choice considered here is the empirical covariance matrix 
\begin{equation*}%\label{eq:J_hat}
\hat C    = \dfrac{1}{n} \sum_{i=1}^n  u(\tilde \theta; Y^{(i)}) u(\tilde \theta; Y^{(i)})^\top,   
\end{equation*}
but  other consistent estimators may be used instead.

Sparse sub-likelihood selection occurs through the minimization of the convex objective (\ref{eq:criterion_empirical}): the $j$th sub-likelihood $\ell_j(\theta)$ is included in the composite likelihood function if $\hat w_j \neq 0$, else $\ell_j(\theta)$ is dropped  and the corresponding parameter estimate is set as $\hat \theta_j = 0$ ($j=1,\dots,p$). The selected composite likelihood function is interpreted as one that maximizes statistical accuracy given a desired level of sparsity. In particular, when $\lambda=0$ the objective $\hat d_0(w)$ corresponds to the so-called finite-sample optimality criterion, a benchmark to find minimum variance estimators for unbiased estimating equations (e.g., see \citealp[Ch. 2]{heyde2008}; \citealp{lindsay2011issues}).  The last  term in (\ref{eq:criterion_empirical}) is a sparsity-inducing penalty discouraging overly complicated composite log-likelihoods.  The geometric properties of the $L_1$-penalty imply that several elements in $\hat w$ are exactly zero for sufficiently  large values of $\lambda$, which also induces sparsity in the estimator $\hat \theta$. A heuristic derivation of  (\ref{eq:criterion_empirical}) is presented in Section \ref{sec:heuristic}.

The considered penalty is adaptive in the sense  that when $\tilde \theta_j$ is near $0$ the $j$th sub-likelihood receives a large penalty.  Adaptive weighting is a  fundamental feature of our method ensuring consistent model selection;  see Theorem \ref{thm:oracle} in  Section \ref{sec:properties}. Our penalty is inspired by the adaptive Lasso penalty introduced  by \cite{zou2006adaptive} in the context of sparse regression; however, the role of our adaptive penalty is completely different because it focuses on the coefficients $w_j$s associated with entire sub-likelihoods, rather than on the parameter elements $\theta_j$s. Penalization on the score space enables one to separate the task of model selection from that of parameter estimation. Hence, differently from existing penalized composite likelihood procedures, the selected estimating equations remain unbiased and  lead to consistent parameter estimators  when the sub-likelihoods are corrected selected.

\subsection{First-order conditions } \label{sec:first_order}

The  empirical objective (\ref{eq:criterion_empirical}) defines a convex minimization 
problem with optimum characterized by the Karush-Kuhn-Tucker (KKT) first-order conditions
\begin{equation}\label{kkt}
\hat C w - \text{diag}(\hat C ) +  \dfrac{\lambda}{n} s = 0\,, 
\end{equation}
where $s=(s_1, \dots, s_p)^\top$ is the sub-gradient of the weighted $L_1$-norm $\sum_{j=1}^p   |w_j|/\tilde \theta^2_j $, that is 
\begin{align*}
s_j =  \frac{1}{ \tilde \theta^{2}_j}  \times 
\left\{
\begin{array}{ccc}
 1 & &\text{if } w_j >0 \\
-1 & &\text{if } w_j <0 \\
\in [ - 1, 1 ] & &\text{if }  w_j = 0
\end{array}
\right. \quad (j=1,\dots, p)\,.
\end{align*}
The solution  can be stated explicitly as $\hat w = (\hat w^\top_{\hat \cA}, 0^\top )^\top$ where
\begin{equation} \label{eq:solution_empirical}
 \hat w_{ \hat \cA}  =    
\hat C_{\hat \cA}^{-1} \left[ \text{diag}(\hat C_{\hat \cA})  -  \dfrac{\lambda}{n}    s_{\hat \cA}\right], 
\end{equation}
and $\hat \cA = \{ j: \hat w_j \neq 0 \}$. Since $\hat \theta_j = \tilde \theta_j I( \hat w_j \neq 0)$, the set $\hat \cA$ coincides with the index set of the estimated nonzero parameters.

Inspection of (\ref{kkt}) provides further insight on the selection process. Consider first the case where $\hat w_j = 0$ for some $j=1,\dots,p$. For this to be true we must have that  
\begin{equation} \label{eq:KKTj}
  \sum_{i=1}^n u_j(\tilde\theta_j; Y^{(i)}) \times {\text{res}}^{(i)}_j =      \lambda    s_j,        
\end{equation}
where ${\text{res}}^{(i)}_j$ is the $i$th pseudo-residual for the $j$th parameter equal to
\begin{equation*}
{\text{res}}^{(i)}_j  = u_j(\tilde \theta_j; Y^{(i)})  -  \sum_{k=1}^p u_k(\tilde \theta_k; Y^{(i)}) \hat w_k. 
\end{equation*}
Taking absolute value on both sides of (\ref{eq:KKTj}) and re-arranging show that a sufficient condition for $\hat w_j \neq 0$ is  
\begin{equation*} %\label{eq:KKTj2}
 Z^2_j  =  \tilde \theta ^2_j  \times \left\vert  \sum_{i=1}^n u_j(\tilde \theta_j; Y^{(i)}) \times \text{res}^{(i)}_j \right\vert > \lambda\,,
\end{equation*}
and $\hat w_j = 0$ if $Z^2_j \le \lambda$. The above inequality reminds us of an acceptance  region for the null hypothesis $\theta_j = 0$, suggesting that $\lambda$ may be selected by considering some appropriate form of error control for multiple tests. For example, if all the hypotheses $H_{0j}: \theta_j = 0$ ($j= 1, \dots, p$) are true, Theorem \ref{thm:oracle} implies $\hat w = o_p(1)$; therefore, each $Z^2_j$ behaves like $ n \tilde \theta ^2_j/SE(\tilde \theta_j)^2$ with $SE(\tilde \theta_j)^2 = \sum_{i=1}^n u_j(\tilde \theta_j; Y^{(i)})^2$ and follows  asyptotically a chi-square distribution with one degree of freedom.

\subsection{Heuristic derivation of the model-selection criterion} \label{sec:heuristic}

Criterion (\ref{eq:criterion_empirical}) used for computing the sparse composition rule comes from the notion of $O_F$-optimality in the theory of unbiased estimating equations \cite[Chapter 2]{heyde2008}. Let $u^{CL}(\theta) = W u(\theta)$ be a  $p \times 1$ composite likelihood score vector where $W$ is some $p\times p$ matrix possibly depending on $\theta$. Fixed-sample optimality prescribes to   minimize the distance between the composite likelihood and the maximum likelihood scores in some appropriate matrix sense. Following \cite{lindsay2011issues}, we consider minimizing the following least squares criterion  over $p \times p$ diagonal matrices $W$:
$$
 E\left[\{u^{ML}(\theta;Y) - W u(\theta; Y) \} \{u^{ML}(\theta;Y) - W u(\theta; Y)\}^\top\right].
$$
When $W$ is a diagonal matrix, the solution of the above least squares objective is a diagonal matrix $W_0(\theta)$ with diagonal elements given by the $p \times 1$ vector
\begin{equation*}%\label{eq:optimal_weights2}
w_0(\theta) = \{C(\theta)\}^{-1} \text{diag}\{ C(\theta)\}\,.
\end{equation*}
%where $C(\theta)= \var\{ u(\theta; Y)\}$; 
When it exists, $w_0(\theta)$ is the minimum of the quadratic objective $d_0(\theta; w) = w^\top C(\theta) w/2  -   w^\top \text{diag}\{C(\theta)\}$.

Although $d_0(w;\theta)$ is a sensible criterion for improving the accuracy of an estimator, it is not helpful for model selection on its own. Without additional information concerning the distance of each $\theta_j$ from zero, it is impossible to discriminate useful parameters and obtain sparsity. Thus, $d_0(\theta; w)$ is augmented by a penalty function, leading to the penalized $O_F$-optimality criterion  
\begin{equation}\label{eq:criterion_ideal}
d_\lambda(w; \theta) = d_0(\theta; w)  + \lambda \sum_{j=1}^p |w_j|/\theta^2_j   =  \dfrac{1}{2} w^\top C(\theta) w  -   w^\top \text{diag}\{C(\theta)\} + \lambda \sum_{j=1}^p |w_j|/\theta^2_j.
\end{equation}
The proposed penalty is justified from a model-selection viewpoint. For any $\lambda>0$, we have
$$ \nonumber
d_\lambda(w;\theta) =
\left\{ 
\begin{array}{lll}
\dfrac{1}{2} w^\top_{\cA} C_{\cA}(\theta) w_{\cA} - w_{\cA}^\top \text{diag}\{ C_{\cA}(\theta)\} + \sum_{j \in \cA}|w_j|/\theta^2_j &  & \text{if } \theta_{j} \neq 0 \ \ \forall j \notin \cA \\
\infty & & \text{otherwise}.
\end{array}
\right.
$$
Since $d_\lambda(w; \theta)$ is convex in $w$, the unique minimum can be expressed as $w(\theta) = (w_{\cA}(\theta)^\top, 0^\top)^\top$  having $p^\ast$ nonzero elements in
\begin{equation*} %\label{eq:population_w}
 w_{ \cA}(\theta)  =    
C_{\cA}(\theta)^{-1} \left[ \text{diag}\{ C_{\cA}(\theta)\}  - \lambda \delta_{\cA}(\theta)\right], 
\end{equation*}
given that $\delta(\theta)$ is the $p \times 1$ vector with components equal to $\text{sign}\{w_{ j}(\theta)\}/ \theta^2_j$ $(j=1,\dots,p)$. Replacing the expectations in (\ref{eq:criterion_ideal}) by sample averages and plugging-in the root-$n$ consistent preliminary estimator $\tilde \theta$ lead to the empirical criterion defined in (\ref{eq:criterion_empirical}).

\subsection{A coordinate-descent algorithm}

The preliminary estimates $\tilde \theta_j$ $(j =1,\dots, p)$ are often easy to find using univariate approaches such as Fisher scoring, with an overall computing cost of order $p$. The convex optimization problem  (\ref{eq:criterion_empirical}) is addressed using a coordinate descent algorithm, whereby minimization is achieved along one direction of $w=(w_1, \dots, w_p)^\top$ at a time. For the vector $a$, let $a_{-j}$ denote the vector $a$ without its $j$th element and use   $\tilde u_j^{(i)}=u_j(\tilde \theta_j; Y^{(i)})$ to express the $j$th score function for the $i$th observation evaluated at   $\theta_j = \tilde \theta_j$ ($j=1,\dots,p$). The $j$th element $w_j$ is updated by solving the scalar equation  
\begin{align*}%\label{eq:partial_deriv}
0 = \dfrac{\partial \hat d_\lambda(w)}{\partial w_j}  = \dfrac{1}{n}\sum_{i=1}^n \tilde u^{(i)}_{j}(\tilde u^{(i)}_{j} w_j +  \tilde u_{-j}^{(i) \top}  w_{-j}  - \tilde u^{(i)}_{j}) + \dfrac{\lambda}{n} \dfrac{\text{sign}(w_j)}{\tilde \theta^2_j}\,.
\end{align*}
%Re-arranging the above equation leads  to the following coordinate descent update. 
Given the value $w^{[s]}=(w^{[s]}_1, \dots w^{[s]}_p)^\top$ at the current step $s$, 
%in one iteration of the algorithm we compute, 
the coordinate descent update for the $j$th component is thus equal to
\begin{align*}
w^{[s+1]}_j 
& =  \dfrac{1}{\sum^n_{i=1} (\tilde u^{(i)}_{j})^2 } S \left\{ \sum^n_{i=1} (\tilde u^{(i)}_{j})^2 - \sum_{i=1}^n \tilde u^{(i)}_j \tilde u_{-j}^{(i)}  w^{[s]}_{-j} ; \ \dfrac{\lambda}{ \tilde \theta^2_j}  \right\} \quad (j=1,\dots,p)\,,
\end{align*}
where $S(x; \lambda)= \text{sign}(x)(|x| - \lambda)_+$  is the soft-thresholding operator with $x_+=\max\{0, x\}$.

\section{Properties for large $n$ and $p$}\label{sec:properties}

We study here the oracle properties of our method guaranteeing that selection of true nonzero parameters occurs with probability going to 1. The number of sub-likelihoods $p=p_n$ is assumed to diverge with the sample size $n$. Although most quantities in this section are functions of $n$, this dependence is left implicit when clear from the context.   Let  $\rho_{\text{min}}$ and $\rho_{\text{max}}$ be the smallest and the largest eigenvalues of $\hat C$, respectively. Denote by $w = w(\theta)$ the optimal sparse composition rule minimizing the population objective in (\ref{eq:criterion_ideal}). Given a vector $v$, $\Vert v\Vert_2$ indicates its Euclidean norm; for a matrix $A$, the induced operator norm is defined as $\Vert A \Vert_2 = \sup_{\Vert x \Vert_2 = 1} \Vert Ax \Vert_2$. 

Using this notation, the following conditions can be stated:

\begin{itemize}
\item[A1] Consistent preliminary estimators: for all $j= 1,\dots ,p$, the preliminary estimator $\tilde \theta_j$ satisfies $|\tilde \theta_j - \theta_j |=o_p(1)$ and $|\tilde \theta_j - \theta_j | = O_p(n^{-1/2})$.
\item[A2]  Consistent pairwise score covariance estimators:  $\Vert \hat C - C(\theta) \Vert_2=O_p(\sqrt{p/n})$.
\item[A3] $k_1 \leq \rho_{\text{min}} \leq \rho_{\text{max}} \leq k_2$, where $k_1$ and $k_2$ are positive constants. 
\item[A4]  $n^{-1} \sum_{i=1}^n   \tilde u^{(i)}_{j}\tilde u_{k}^{(i)} = O_p(1)$ for all $j \in \cA$ and $k \notin \cA$.  
\end{itemize}

\begin{theorem}[Consistency] \label{thm:consistency}
Under conditions A1--A3, if $\lambda n^{-1/2}\rightarrow 0$, then we have
$$
\left\Vert\hat w - w \right\Vert_2 = O_p\left( \sqrt{\frac{p}{n}} \right).
$$
\end{theorem}

The above theorem states that the empirical composition rule $\hat w$ is a root-($n/p$) consistent estimator of the optimal sparse composition rule $w$, which contains zero elements corresponding to irrelevant estimating equations;  the nonzero elements of $w$ represent an optimal composition rule on the set $\cA$.

\begin{theorem}\label{thm:sparse_solution}
Define 
\begin{equation}\label{eq:sparse_solution}
\hat w_{\cA} = \underset{w_{\cA}}{\arg\min} \left\{ \dfrac{1}{2}w_{\cA}^\top \hat C_{\cA}    
w_{\cA} -  w_{\cA}^\top \mathrm{diag} (\hat C_{\cA})  +    \dfrac{\lambda}{n}  \sum_{j \in \cA}  \dfrac{     |w_{j}|}{\tilde \theta^2_j} \right\} \,. 
\end{equation}
Under conditions A1--A4, if $\lambda$ satisfies $\lambda n^{-1/2} \rightarrow 0$ and $\lambda n^{1/2} \rightarrow \infty$, then with probability tending to 1, $(\hat w_{\cA}^\top, 0^\top )^\top$ is the minimizer of the criterion $\hat d_\lambda(w)$ in (\ref{eq:criterion_empirical}).
\end{theorem}

This theorem provides an asymptotic description of the solution of our selection criterion. With probability tending to 1, the empirical composition rule for the scores corresponding to irrelevant parameters is zero.  This result also suggests that in high-dimensions our selection rule should enjoy model-selection consistency, which is stated in the next theorem.

\begin{theorem}[Model-selection consistency] \label{thm:oracle}
Under conditions A1--A4, if $\lambda$ satisfies $\lambda n^{-1/2} \rightarrow 0$ and $\lambda n^{1/2} \rightarrow \infty$, then
$P( \hat\calA   = \cA) \rightarrow 1$, as $n \rightarrow \infty$.
\end{theorem}

The above result states that our empirical composition rule is consistent for model selection, i.e. the nonzero $\hat w_j$ correspond to relevant parameters as the number of parameters grows with the sample size. A direct consequence of this is the normality of the selected parameter estimates. Under suitable regularity conditions, %\Ccom{REF?}, 
$n^{1/2} ( \tilde \theta - \theta )_{ \cA}$ follows  a normal distribution with asymptotic variance $G^{-1}_{\cA}(\theta)$,  where 
\begin{equation}
 G_{ \cA}(\theta) =  H_{\cA}(\theta)^{-1} C_{\cA}(\theta) H_{\cA}(\theta)^{-1} 
 \end{equation}
 is the $p^{\ast}\times p^\ast$  matrix with components
 \begin{align}
  H_{\cA}(\theta) =   E \left\{ \nabla u_{\cA}(\theta;Y)\right\}, \ \ 
  C_{\cA}(\theta) =   E \left\{ u_{\cA}(\theta;Y) u_{\cA}(\theta;Y)^{\top} \right\}.
\end{align}
Since $\hat \theta_j = \tilde \theta_j I(\hat w_j \neq 0)$, Theorem  \ref{thm:oracle} and Slutsky's theorem imply that $n^{1/2} G^{1/2}_{\hat \cA}(\theta) ( \hat\theta - \theta )_{ \hat \cA}$ converges in distribution to a $p^\ast$-variate normal random variable with zero mean and identity covariance matrix.

\section{Examples}

\subsection{Sparse multivariate location} \label{sec:Example1}

Sparse multivariate location estimation is central for a number of statistical analyses, including  variable screening and multiple hypothesis testing, and represents the base for more complicated setups. Considering samples from $Y \sim N_p(\theta, \Sigma)$,  the $j$th marginal score is $u_j(\theta_j; y) = (y_j -\theta_j)/\sigma_j^2$ and yields  $\tilde \theta_j = \bar Y_j =n^{-1} \sum_{i=1}^n Y_j^{(i)}$ ($j=1,\dots,p$). The $p\times p$ empirical score covariance matrix is   
$
\hat C= n^{-1}\sum_{i=1}^n  (Y^{(i)}- \bar Y_j)(Y^{(i)}-\bar Y_j)^\top  
$ and $\hat w$ is found as in (\ref{eq:solution_empirical}). When $\Sigma= I_p$, with $I_p$ denoting the identity matrix of order $p$, we have the explicit solution
$$
\hat w_j  =  \left( 1 - \dfrac{\lambda \hat C_{jj} }{ n \bar Y^{2}_j }  \right)I\left(\dfrac{n \bar Y_j^2}{\hat C_{jj}} > \ \lambda  \right) \quad   (j =1,\dots, p)\,,
$$
where $\hat C_{jj}$ indicates the $(j,j)$th entry of $\hat C$.
The final sparse estimator has $j$th component equal to
$
\hat \theta_j   = \bar Y_j I ( \hat w_j \neq 0  ) = \bar Y_j \  I (n \bar Y_j^2/\hat C_{jj} > \lambda   )$ $(j =1,\dots, p)$. If $j \in \cA$,   $n\bar Y_j^2/\hat C_{jj}^2$ diverges with $n$, which implies the oracle property  $P(j \in \hat \cA ) \rightarrow 1$. If $j \notin \cA$, the quantity  ${n \bar Y_j^2}/{\hat C_{jj}^2}$ typically converges in distribution to a chi-square random variable with one degree of freedom. A slowly diverging $\lambda$, e.g. $\lambda = O(\log n)$ would suffice to control the Type I error probability, i.e. the probability of selecting $\theta_j$ when $\theta_j=0$.

A useful extension is that of  multivariate generalized linear models where each $Y_j$ $(j=1,\dots,p)$ is assumed to depend on a predictor $x$ through
$
\mu_j   =  E(Y_j)  = g^{-1}(\alpha+\theta_j x )
$, 
for some invertible link function $g$. 
For observations $(Y^{(i)}, x^{(i)})$ $(i =1,\dots, n)$, the marginal preliminary estimator $\tilde \theta_j$ is found by solving estimating equations $\sum_{i=1}^n u_j(\theta_j; Y^{(i)}) = 0$. With canonical links, the scores are $u_j(\theta_j; y) = (y_j - \mu_{j}) \partial \mu_{j}/\partial \theta_j$ and the $(j,k)$th entry in $\hat C$ is 
 $$
 \hat C_{jk} =  \dfrac{1}{n}\sum_{i=1}^n (Y_j^{(i)} - \tilde\mu^{(i)}_{j} ) (Y_k^{(i)} - \tilde\mu^{(i)}_{k} ) \left( \dfrac{ \partial \tilde \mu^{(i)}_{j}}{\partial \tilde \theta_j} \right)^\top  \left( \dfrac{ \partial \tilde\mu^{(i)}_{j}}{\partial \tilde\theta_k} \right)   \quad (j,k=1,\dots, p),
$$
where $\tilde\mu^{(i)}_{j} = g^{-1}(\alpha+\tilde\theta_j x^{(i)} )$. When the scores are uncorrelated, the final estimator can be written explicitly as 
$
\hat \theta_j  =  \tilde \theta_j \  I(n \tilde \theta_j^2 \hat C^{-1}_{jj} > \lambda )$.

\subsection{Sparse correlation graphs} \label{sec:Example2}
Sparse covariance and correlation matrix estimation is a fundamental problem in statistics. A variety of strategies have been proposed for reducing the number of parameters in large covariance matrices. Among those are penalized likelihood methods \citep{bien11, rothman12} and thresholding methods \citep{el08, rothman09, cai11}. Let $Y \sim N_d(0,R)$, where $R=R(\theta)$ is a sparse correlation matrix  with $(j_1,j_2)$th entry denoted by $\{R\}_{j_{1}j_{2}}$; thus for $j=1,\dots,p$ we have that $\theta_j= \{R\}_{j_{1}j_{2}}$ with $1\leq j_{1}<j_{2} \leq d $.
Since marginal univariate sub-likelihoods do not contain information on $\theta$, we consider unit pairwise sub-likelihoods obtained by taking $p = d(d-1)/2$ bivariate normal log-densities for the pairs $(y_{j_1}, y_{j_2})$.  
%\Ccom{OR:
%\begin{equation}
%		\log f_{j}(y_{j_1}, y_{j_2}; \theta_{j}) =  - \frac{1}{2} \log (1-\theta_{j}^2) -  \frac{(y^{(i)}_{j_1})^2 
%			+(y^{(i)}_{j_2})^2 }{2(1-\theta_{j}^2)} + \frac{\theta_{j} y_{j_1}^{(i)} y_{j_2}^{(i)}}{1-\theta_{j}^2} + \text{const}\,, 
%\end{equation}}
Each corresponding $j$th score
%1\leq j_{1}<j_{2} \leq d 
equals then
\begin{equation}\label{eq:pairwise_scores}
		u_{j}(\theta_{j};  y_{j_1}, y_{j_2}) = 
		(1+ \theta_{j}^2) y_{j_1} y_{j_2}   - \theta_{j}( y_{j_1}^2 
		+y_{j_2}^2 )
		+   \theta_{j} ( 1-\theta_{j}^2)\,.
\end{equation}
 
With multivariate ordered categorical data, sparse correlation matrices may be obtained using our method in combination with a latent variable model. Specifically, if $Y_{j}\in \{0,1, \dots,$ $L+1\}$ as in \cite{han2012composite}, one may consider a latent  vector $Z = (Z_{1}, \dots, Z_{d}) \sim N_d(0,
R)$, where $R= R(\theta)$ is defined as above. 
Pairwise scores indexed by $j = (j_1,j_2)$ are 
%with $1 \leq j_1<j_2 \leq d$ are
\begin{align*}
	u_{j}(\theta_{j}; y_{j_1}, y_{j_2}) & =   \dfrac{\partial}{\partial \theta_{j}}\log P(Y_{j_1} = y_{j_1}, Y_{j_2}= y_{j_2}; \theta_{j}) \\
	& =  \dfrac{\partial}{\partial \theta_{j}} \log \int_{\Gamma_{j_1}} \int_{\Gamma_{j_2}}   \phi(z_1, z_2; \theta_{j}) dz_1 dz_2\quad (j=1,\dots,p; 1 \leq j_1<j_2 \leq d)\,,
\end{align*}
where $\phi(\cdot, \cdot; \theta_{j})$ is the bivariate normal density with zero mean, unit variances and correlation equal to $\theta_{j}$,  $\Gamma_{j_1}$ and $\Gamma_{j_2}$ are intervals in $\{(-\infty, \gamma_1], (\gamma_1, \gamma_2), \dots, (\gamma_L, \infty)\}$ containing $y_{j_1}$ and $y_{j_2}$, respectively, with  $\gamma_1<\gamma_2< \dots< \gamma_L$ fixed thresholds.

\section{Numerical studies}

\subsection{Monte Carlo simulations} \label{sec:simu}

The model-selection and  estimation properties of the proposed method  are illustrated through three Monte Carlo experiments. All the results are based on 2500 samples of size $n=250$, enabling comparisons with the maximum likelihood benchmark.

\noindent {\it Setting 1:} Sparse location estimation in the $p$-variate normal model $Y \sim N_{p}(\theta, \Sigma)$ with $p=100$. The mean vector %$\theta=(\theta_{\cA},0, \dots,0)^\top$ where 
$\theta$ has $p^\ast = 25$ nonzero elements
$\theta_{\cA}=(5,\dots 5, 4,\dots, 4,$ $3, \dots, 3, 2,\dots 2,$ $ 1,\dots 1)^\top$ and $\Sigma$ is such that $\{\Sigma\}_{jj} = 1 $ for all $j$ and $\{\Sigma\}_{jk}  \in \{0,0.5\} $ ($j \neq k$). {\it Setting 2:}  Sparse location estimation in the $p$-variate probit regression with $p=100$. The $j$th binary response
$Y_{j}= I(Z_j \geq 0)$ ($ j =1,\dots, p$) is generated based on $Z \sim N_p(\mu, \Sigma)$ where $\mu_j= 0.1 + \theta_jx$, the nonzero $p^*=25$ probit coefficients are $\theta_{\cA}=(1.5,\dots, 1.5, 1.25,\dots 1.23, 1,\dots 1,0.75,$ $\dots 0.75, 0.5, \dots, 0.5)^\top$, $\Sigma$ is as in Setting 1, and the covariate $x$ is independently drawn from a $N(0,1)$. {\it Setting 3:} Sparse correlation estimation in the $d$-variate normal model $Y \sim N_{d}(0,R)$ with $d=15$.
The $p =d(d-1)/2=105$ parameters in the correlation matrix $R=R(\theta)$ correspond to entries $\theta_j= \{R \}_{j_1j_2}$ ($ j =1,\dots, p$; $1\leq j_{1}<j_{2} \leq d $), among which $p^\ast = 10$ are nonzero; we consider for such values either uniform correlations $\theta_{j}  = 0.5$ or the Toeplitz structure  $\theta_{j}  = e^{-0.1|j_1-j_2|}$.
 In Settings 1 and 2, the score covariance matrix $\hat C$ for the objective (\ref{eq:criterion_empirical}) is obtained based on the marginal scores reported in Section \ref{sec:Example1}, whereas the pairwise scores described in  Section \ref{sec:Example2} are used for Setting 3.

Tables \ref{tab1}, \ref{tab2} and \ref{tab3}  show  estimates for the true positive probability, true negative  probability  and  false discovery probability computed as 
\begin{gather*}
\mathrm{TPP}= 
 \dfrac{\#\{j : \hat \theta_{j} \neq 0, \theta_j \neq 0\} }{p^*},  \  \mathrm{TNP}=% 
 \dfrac{\#\{j : \hat \theta_{j}= 0, \theta_j = 0\} }{p-p^*}, \
\mathrm{FDP}=  
\dfrac{\#\{j : \hat \theta_{j} \neq 0, \theta_j = 0\} }{\hat p^*},
\end{gather*}
along with the mean number of selected parameters, for equally spaced values of $\lambda$ on the log-scale. The sparse combination of likelihood scores is found to be reliable for model selection, especially under the first framework based on normal data, thus confirming our theoretical results in Section \ref{sec:properties}. For judiciously chosen values of  $\lambda$, the procedure exhibits remarkable properties both in terms of type I error (FDP) and of power (TPP). As expected, the performance appears to be affected by the size of correlation among the score components.

\begin{figure}[h!]
\begin{center}
	\includegraphics[height=5.7cm]{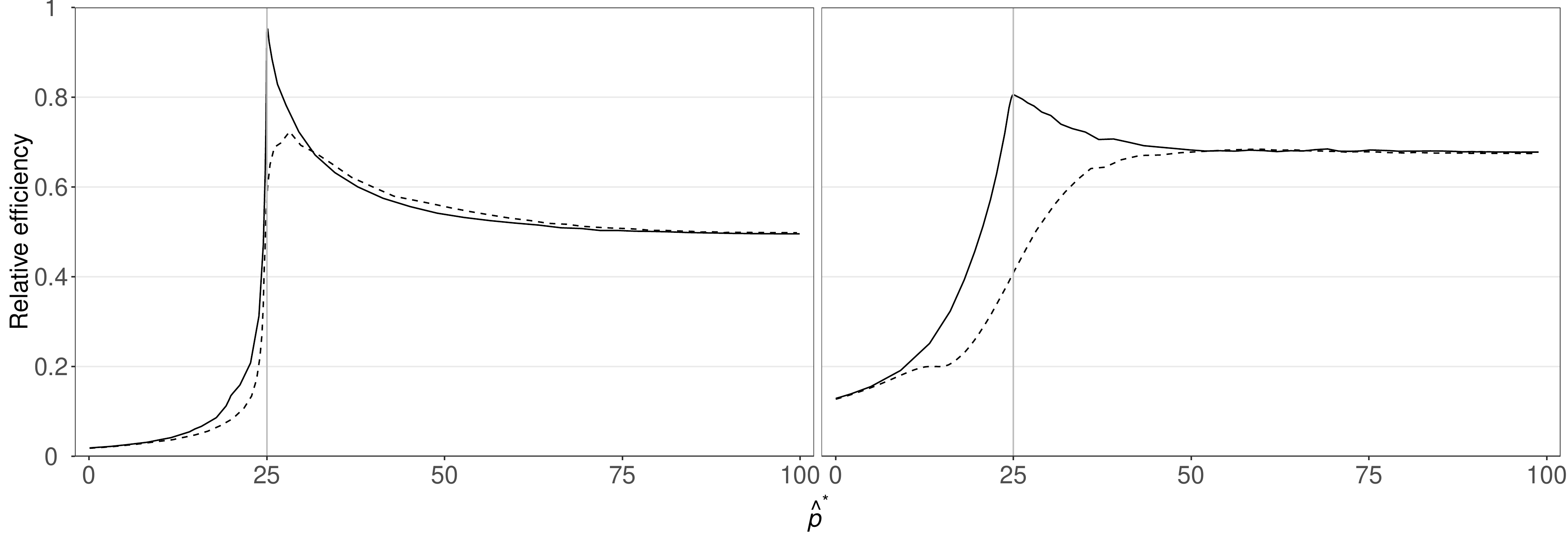}
	\caption{Monte Carlo estimates for the relative efficiency of the sparse composite likelihood estimator with respect to the oracle maximum likelihood estimator versus the average number of selected parameters $\hat p^\ast$. The trajectories for the multivariate normal location (left panel) and multivariate probit regression (right panel) correspond to uncorrelated scores (solid) and correlated scores (dashed). The vertical line at $p^*=25$ indicates the true number of nonzero parameters.}
	\label{figRE}
\end{center}
\end{figure}

Figure \ref{figRE} compares the efficiency of the sparse composite likelihood estimator $\hat\theta$ with that of the oracle maximum likelihood estimator $\hat\theta_{\text{or:mle}}$. The latter corresponds to the maximum likelihood estimator for $\theta_{\cA}$, while the remaining parameters are set equal to zero. This is clearly an unattainable benchmark since it assumes perfect knowledge of the underlying model structure. The curves in the graph show Monte Carlo estimates of the relative efficiency, computed as the ratio of the  root mean squared error for the two estimators, $RMSE(\hat\theta_{\text{or:mle}})/  RMSE (\hat\theta)$, against the estimated average of selected parameters. For uncorrelated  scores, the relative efficiency is largest at $p^\ast$, and particularly close to one in Setting 1. If the scores are all correlated with $\rho=0.5$,   estimation accuracy is hindered by the less reliable selection as shown in Tables \ref{tab1} and \ref{tab2}.
Especially in the multivariate probit setting, the maximum efficiency is reached after $p^*$, yet remains reasonably high around 0.7.

\begin{table}[h]
	\caption{Monte Carlo estimates for the number of selected components, percent true positive probability (TPP), percent true negative probability (TNP) and percent false discovery probability (FDP) for  the  $p$-variate location  model $Y \sim N_{p}(\theta, \Sigma)$ with $p=100$, $p^\ast=25$ nonzero elements in $\theta$, and matrix $\Sigma$ with $\{ \Sigma \}_{jj}=1$ and off-diagonal elements $\{ \Sigma \}_{jk}$ ($j\neq k$). Results are based on 2500 Monte Carlo samples of size $n=250$.}
	\label{tab1}
	\centering
	\begin{tabular}{c|cccc|cccc}
		& \multicolumn{4}{c}{$\{ \Sigma \}_{jk}=0$} & \multicolumn{4}{c}{$\{ \Sigma \}_{jk}=0.5$} \\ 
		$\lambda$   & $\hat p^*$ & TPP & TNP & FDP & $\hat p^*$ & TPP & TNP& FDP \\ 
		\toprule
		~~0.750 & 52.157 & 100.0 & ~63.8 & 51.8 & 47.900 & 100.0 & ~69.5 & 45.4 \\ 
		~~1.292 & 43.231 & 100.0 & ~75.7 & 41.7 & 40.334 & ~99.9 & ~79.5 & 34.6 \\ 
		~~2.225 & 35.057 & 100.0 & ~86.6 & 28.2 & 32.801 & ~99.9 & ~89.6 & 19.3 \\ 
		~~3.832 & 29.133 & 100.0 & ~94.5 & 13.8 & 27.846 & ~99.8 & ~96.1 & ~7.7 \\ 
		~~6.599 & 26.038 & 100.0 & ~98.6 & ~3.9 & 25.475 & ~99.5 & ~99.2 & ~1.8 \\ 
		~11.365 & 25.130 & 100.0 & ~99.8 & ~0.5 & 24.793 & ~99.0 & ~99.9 & ~0.2 \\ 
		~19.574 & 24.994 & ~99.9 & 100.0 & ~0.0 & 24.448 & ~97.8 & 100.0 & ~0.0 \\ 
		~33.713 & 24.973 & ~99.9 & 100.0 & ~0.0 & 23.520 & ~94.1 & 100.0 & ~0.0 \\ 
		~58.062 & 24.950 & ~99.8 & 100.0 & ~0.0 & 20.664 & ~82.7 & 100.0 & ~0.0 \\ 
		100.000 & 24.846 & ~99.4 & 100.0 & ~0.0 & 15.937 & ~63.7 & 100.0 & ~0.0 \\
		\bottomrule
	\end{tabular}
\end{table}

\begin{table}[h!]
	\caption{Monte Carlo estimates for the number of selected components, percent true positive probability (TPP), percent true negative probability (TNP) and {percent} false discovery probability (FDP)  for the coefficients of the $p$-variate probit regression $Y_{j}=I(Z_j \geq 0)$, given $Z\sim N_{p}(\mu, \Sigma)$ where $\mu_j= 0.1 + \theta_jx$ ($j=1,\dots, p$), with $p=100$ and $p^*=25$ nonzero $\theta_j$s. The matrix $\Sigma$ has entries $\{ \Sigma \}_{jj}=1$ and $\{ \Sigma \}_{jk}$ ($j\neq k$). Results are based on 2500 Monte Carlo samples of size $n=250$.}
	\label{tab2}
	\centering
	\begin{tabular}{c|cccc|cccc}
		\multicolumn{1}{c}{} & \multicolumn{4}{c}{$\{ \Sigma \}_{jk}=0$} & \multicolumn{4}{c}{$\{ \Sigma \}_{jk}=0.5$} \\ 
		$\lambda$    & $\hat p^*$ & TPP & TNP & FDP & $\hat p^*$ & TPP & TNP & FDP \\ 
		\toprule
		~0.200 & 69.930 & 99.9 & ~40.0 & 64.2 & 53.053 & 99.4 & ~62.4 & 51.0 \\ 
		~0.360 & 62.330 & 99.7 & ~50.1 & 59.8 & 40.558 & 98.5 & ~78.8 & 35.4 \\ 
		~0.649 & 53.742 & 99.5 & ~61.5 & 53.5 & 32.046 & 96.4 & ~89.4 & 21.9 \\ 
		~1.170 & 44.482 & 99.4 & ~73.8 & 43.7 & 26.484 & 92.4 & ~95.5 & 11.4 \\ 
		~2.107 & 35.487 & 99.4 & ~85.8 & 29.5 & 22.766 & 86.6 & ~98.5 & ~4.4 \\ 
		~3.796 & 28.839 & 99.3 & ~94.6 & 13.5 & 19.956 & 78.9 & ~99.7 & ~0.9 \\ 
		~6.840 & 25.565 & 99.0 & ~98.9 & ~3.0 & 17.354 & 69.2 & ~99.9 & ~0.1 \\ 
		12.323 & 24.597 & 98.1 & ~99.9 & ~0.2 & 15.117 & 60.4 & 100.0 & ~0.0 \\ 
		22.202 & 23.264 & 93.1 & 100.0 & ~0.0 & 13.096 & 52.3 & 100.0 & ~0.0 \\ 
		40.000 & 19.006 & 76.0 & 100.0 & ~0.0 & ~9.660 & 38.5 & 100.0 & ~0.0 \\ 
		
		\bottomrule
	\end{tabular}
\end{table}

\begin{table}[h!]
	\caption{Monte Carlo estimates for the number of selected components, percent true positive probability (TPP), {percent} true negative probability (TNP) and {percent} false discovery probability (FDP)  for the $p=105$ correlations $\theta_j=\{R \}_{j_1j_2}$ ($ j =1,\dots, p$; $1\leq j_{1}<j_{2} \leq d $) in the model  $Y \sim N_{d}(0,R)$ with $d=15$. The  $p^\ast = 10$ nonzero elements equal either $\theta_{j}=0.5$ or $\theta_{j}=e^{-0.1|j_1-j_2|}$. Results are based on 2500 Monte Carlo samples of size $n=250$.
}
	\label{tab3}
	\centering
	\begin{tabular}{c|cccc|cccc}
		\multicolumn{1}{c}{} & \multicolumn{4}{c}{$\theta_{j}=0.5$} & \multicolumn{4}{c}{$\theta_{j}=e^{-0.1|j_1-j_2|}$} \\ 
		$\lambda$    & $\hat p^*$ & TPP & TNP & FDP& $\hat p^*$ & TPP & TNP & FDP \\ 
		\toprule
			0.300 & 31.468 & 93.3 & ~76.7 & 69.9 & 28.607 & 99.1 & ~80.3 & 64.7 \\ 
		0.426 & 24.960 & 91.3 & ~83.3 & 62.7 & 23.041 & 98.4 & ~86.1 & 56.3 \\ 
		0.604 & 19.025 & 89.8 & ~89.4 & 51.7 & 17.990 & 97.1 & ~91.3 & 44.8 \\ 
		0.857 & 14.494 & 90.1 & ~94.2 & 36.4 & 13.997 & 94.9 & ~95.3 & 30.7 \\ 
		1.216 & 11.494 & 90.2 & ~97.4 & 20.2 & 11.218 & 92.0 & ~97.9 & 16.8 \\ 
		1.726 & ~9.772 & 89.5 & ~99.1 & ~7.7 & ~9.428 & 87.5 & ~99.3 & ~6.6 \\ 
		2.450 & ~8.965 & 87.7 & ~99.8 & ~2.0 & ~8.214 & 80.6 & ~99.8 & ~1.8 \\ 
		3.476 & ~8.453 & 84.2 & 100.0 & ~0.3 & ~7.090 & 70.6 & 100.0 & ~0.4 \\ 
		4.933 & ~7.836 & 78.3 & 100.0 & ~0.0 & ~5.778 & 57.7 & 100.0 & ~0.1 \\ 
		7.000 & ~6.948 & 69.5 & 100.0 & ~0.0 & ~4.637 & 46.4 & 100.0 & ~0.0 \\
		\bottomrule
	\end{tabular}
\end{table}

\subsection{Analysis of the cell signaling data}

\begin{figure}
\begin{tikzpicture}[<->, rotate=60, >=stealth', scale=0.92, font=\footnotesize]
	\tikzstyle{every edge}=[thick, draw, color=black];
	%\tikzstyle{every node}=[draw, color=black];
	\node at ({1*360/11}:2.5cm) (1) {Raf};
	\node at ({2*360/11}:2.5cm) (11) {Jnk};
	\node at ({3*360/11}:2.5cm) (10) {P38};
	\node at ({4*360/11}:2.5cm) (9) {PKC};
	\node at ({5*360/11}:2.5cm) (8) {PKA};
	\node at ({6*360/11}:2.5cm) (7) {Akt};
	\node at ({7*360/11}:2.5cm) (6) {Erk};
	\node at ({8*360/11}:2.5cm) (5) {PIP3};
	\node at ({9*360/11}:2.5cm) (4) {PIP2};
	\node at ({10*360/11}:2.5cm) (3) {Plcg};
	\node at ({11*360/11}:2.5cm) (2) {Mek};
	\path 
	(1) edge (2)
	(2) edge (3)
	(2) edge (4)
	(2) edge (7)
	(2) edge (10)
	
	(3) edge (4)
	(3) edge (7)
	(3) edge (8)
	(3) edge (9)
	(3) edge (10)
	(3) edge (11)
	
	(4) edge (5)
	(4) edge (7)
	(4) edge (9)
	(4) edge (10)
	(4) edge (11)
	
	(6) edge (7)
	(6) edge (8)
	(6) edge (11)
	
	(7) edge (9)
	(7) edge (10)
	(7) edge (11)
	
	(9) edge (10)
	(9) edge (11)
	
	(10) edge (11);
%	\node[above, font=\large\bfseries] at (current bounding box.north) {SCEE};
\end{tikzpicture}\begin{tikzpicture}[<->, rotate=60, >=stealth', scale=0.92, font=\footnotesize]%\textbf{SCEE}
	\tikzstyle{every edge}=[thick, draw, color=black];
	%\tikzstyle{every node}=[draw, color=black];
	\node at ({1*360/11}:2.5cm) (1) {Raf};
	\node at ({2*360/11}:2.5cm) (11) {Jnk};
	\node at ({3*360/11}:2.5cm) (10) {P38};
	\node at ({4*360/11}:2.5cm) (9) {PKC};
	\node at ({5*360/11}:2.5cm) (8) {PKA};
	\node at ({6*360/11}:2.5cm) (7) {Akt};
	\node at ({7*360/11}:2.5cm) (6) {Erk};
	\node at ({8*360/11}:2.5cm) (5) {PIP3};
	\node at ({9*360/11}:2.5cm) (4) {PIP2};
	\node at ({10*360/11}:2.5cm) (3) {Plcg};
	\node at ({11*360/11}:2.5cm) (2) {Mek};
	\path 
	(1) edge (2)
	
	(3) edge (4)
	(3) edge (7)
	(3) edge (10)
	(3) edge (11)
	
	(4) edge (7)
	(4) edge (10)
	
	(6) edge (7)
	
	(7) edge (11)
	
	(9) edge (10)
	(9) edge (11)
	
	(10) edge (11);
	%\node[above, font=\large\bfseries] at (current bounding box.north) {SCEE};
\end{tikzpicture}\begin{tikzpicture}[<->, rotate=60, >=stealth', scale=0.92, font=\footnotesize]%\textbf{SCEE}
	\tikzstyle{every edge}=[thick, draw, color=black];
	%\tikzstyle{every node}=[draw, color=black];
	\node at ({1*360/11}:2.5cm) (1) {Raf};
	\node at ({2*360/11}:2.5cm) (11) {Jnk};
	\node at ({3*360/11}:2.5cm) (10) {P38};
	\node at ({4*360/11}:2.5cm) (9) {PKC};
	\node at ({5*360/11}:2.5cm) (8) {PKA};
	\node at ({6*360/11}:2.5cm) (7) {Akt};
	\node at ({7*360/11}:2.5cm) (6) {Erk};
	\node at ({8*360/11}:2.5cm) (5) {PIP3};
	\node at ({9*360/11}:2.5cm) (4) {PIP2};
	\node at ({10*360/11}:2.5cm) (3) {Plcg};
	\node at ({11*360/11}:2.5cm) (2) {Mek};
	\path 	
	(3) edge (4)
	(3) edge (7)
	
	(6) edge (7)
	
	(9) edge (10)
	(9) edge (11)
	
	(10) edge (11);
%	\node[above, font=\large\bfseries] at (current bounding box.north) {SCEE};
\end{tikzpicture}

\caption{Covariance graphs resulting by selection of pairwise likelihoods based on different sparsity levels: $\hat p^*=25$ (left), $\hat p^*=12$ (middle), $\hat p^*=6$ (right).}
\label{fig:example_correlation}
\end{figure}
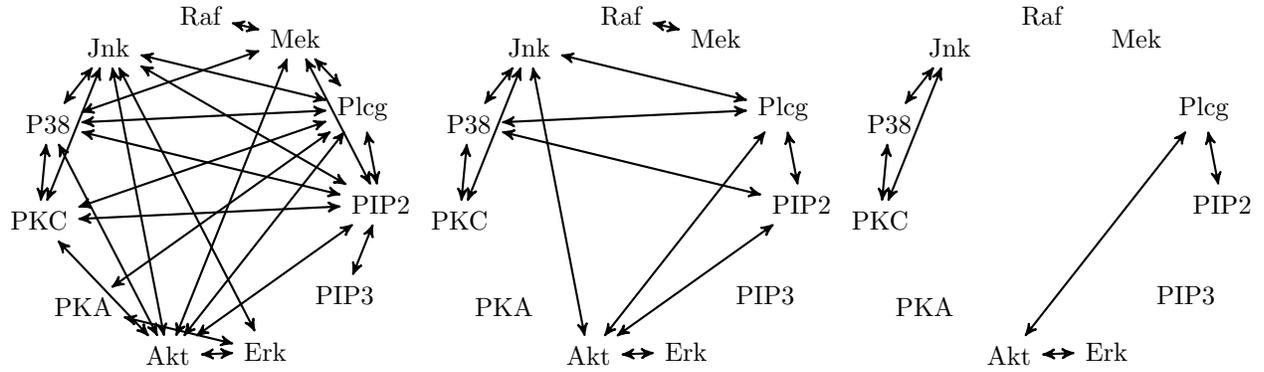

The proposed approach is illustrated through the sparse correlation matrix estimation for the cell signaling dataset.
%, often used in the literature a benchmark dataset. 
The data  consist of flow cytometry measurements of the concentration of $d=11$ proteins in $n =7466$ cells \citep{sachs05}. The sample size is much larger than the total number of correlation parameters $p = d(d - 1)/2 = 55$, which allows us to compare the accuracy of our method to the maximum likelihood benchmark. After standardising the data, our method is implemented using bivariate normal sub-likelihoods for each protein pair with pairwise scores as in (\ref{eq:pairwise_scores}). For illustration purposes, values of $\lambda$ corresponding to 25, 12 and 6 selected correlation parameters are considered as in \cite{bien11}. Figure \ref{fig:example_correlation} shows the corresponding covariance graphs where edges represent nonzero correlations between protein pairs.

Table \ref{tab4} shows the protein pairs corresponding to the six most significant Wald-type maximum likelihood statistics for testing $\text{cor}(Y_{j_1}, Y_{j_2})=0$ $(j_1 < j_2)$ and report the selection made by our strategy for $\hat p^\ast=6$. For comparison, the selections delivered by $L_1$-penalized maximum likelihood and soft-thresholding methods for sparse covariance estimation are also shown; 
%\citep{rothman09, bien11}; 
these are implemented in the R \citep{R} packages  \texttt{spcov} \citep{spcov} and  \texttt{FinCovRegularization} \citep{yan16}, respectively. The pairs resulting from composite likelihood selection are clearly the ones that agree most with the maximum likelihood ranking, with  the only exception being   the pair (Raf, Mek), which is however included when we consider $\hat p^\ast = 12$, as visible in  Figure \ref{fig:example_correlation}.

In order to assess the estimation accuracy, we split the sample in 30 random subsets of approximately 250 observations. We estimate the correlation matrix on each subset at different sparsity levels, and then calculate the root mean squared error over the 30 sub-samples using the maximum likelihood estimator from the whole dataset as the true parameter value.  
By applying our method, we obtain values equal to  $0.191$, $0.247$ and $0.247$ corresponding to selected components $\hat p^\ast = 25$, $12$ and $6$, respectively. For the $L_1$-penalized maximum likelihood the root mean squared error is  and $0.227$, $0.292$ and $0.324$ while for soft-thresholding is  $0.262$, $0.289$ and $0.323$.

%\begin{table}[h!]
%	\caption{}
%	\label{tab3}
%	\centering
%	\begin{tabular}{c|c|c|c|c}
%		Ranking & Edge & SCEE   & $\ell_1$-penalized ML & Soft-thresholding \\ 
%		\hline
%		1 & (Raf, Mek) & $\checkmark$  &  & $\checkmark$ $\checkmark$  \\ 
%		2 & (Plcg, PIP2) & $\checkmark$$\checkmark$ & $\checkmark$ & $\checkmark$ $\checkmark$ \\ 
%		3 & (PKC, P38) & $\checkmark$$\checkmark$ & $\checkmark$$\checkmark$ & $\checkmark$  \\ 
%		4 & (PKC, Jnk) & $\checkmark$$\checkmark$  & $\checkmark$$\checkmark$ &  \\ 
%		5 & (P38, Jnk) & $\checkmark$$\checkmark$  &  & $\checkmark$$\checkmark$ \\ 
%		6 & (Erk, Akt) & $\checkmark$$\checkmark$ & $\checkmark$$\checkmark$  &   \\ 
%		7 & (PKA, P38) &  &  & $\checkmark$$\checkmark$  \\ 
%		8 & (PKA, PKC) &  &  &   \\ 
%		9 & (PKA, Jnk) &  &  &   \\ 
%		10 & (PIP2, PKA) &  &  & $\checkmark$  \\ 
%		11 & (PIP2, P38) & $\checkmark$ &  & $\checkmark$$\checkmark$  \\ 
%		12 & (Akt, Jnk) & $\checkmark$ & $\checkmark$$\checkmark$ &   \\ 
%		
%		\hline
%	\end{tabular}
%\end{table}

\begin{table}[h!]
	\caption{Edges ranked according to the significance of the maximum likelihood Wald-type statistic for the corresponding correlation coefficient, from highest to lowest. The ticks indicate the edges selected by sparse composite likelihood (SCL), maximum likelihood with $L_1$ penalty ($L_1$-ML) and soft-thresholding (ST) methods with $\hat p^*=6$.}
	\label{tab4}
	\centering
	\begin{tabular}{c|c|c|c|c}
		Ranking & Edge & SCL  & $L_1$-ML & ST \\ 
		\hline
		1 & (Raf, Mek) &   & $\checkmark$ & $\checkmark$  \\ 
		2 & (PKC, P38) & $\checkmark$ &  &   \\ 
		3& (Plcg, PIP2) & $\checkmark$ & $\checkmark$ &  $\checkmark$ \\ 
		4 & (PKC, Jnk) & $\checkmark$  &  &  \\ 
		5 & (P38, Jnk) & $\checkmark$  & $\checkmark$ & $\checkmark$ \\ 
		6 & (Erk, Akt) & $\checkmark$ &   &   \\  
		\hline
	\end{tabular}
\end{table}

\section{Discussion}

A selection method to construct sparse composite likelihood functions by maximizing the statistical efficiency of the resulting composite likelihood estimator for a given level of sparsity was introduced.  The nature of our approach is different  from classic composite likelihood penalized strategies \citep{bradic2011penalized, xue2012nonconcave,  gao2017data} since sparsity originates by penalizing entire sub-likelihood scores rather than parameter elements. Conditionally on correct sub-likelihood selection, this feature enables one to retain the unbiasedness of the original sub-likelihood equations and the consequent first-order properties of the composite likelihood estimator. Under the conditions in Section \ref{sec:properties}, the proposed method fulfills model-selection consistency as $p$ diverges with  $n$, implying that all the nonzero parameters are selected with probability going to 1 and the final estimator retains the asymptotic distribution of the composite likelihood estimator on the relevant parameters. This is confirmed by our numerical studies showing reliable model selection and accurate parameter estimation for sparse location and correlation models. 

When sub-likelihoods are functions of a relatively small number of parameters the selection method is extremely simple to implement and fast in execution, which is beneficial for the analysis of large-$p$ problems. A straightforward generalization retaining these computational advantages is obtained by constructing scores and differentiating the composite likelihood in the usual way, i.e. $u_j(\theta; y) = \partial \ell(\theta; y)/\partial \theta_j$ $(j=1,\dots, p)$, when each sub-likelihood has a finite, and possibly small, number of parameters; this setting includes the traditional  composite likelihood framework described  in \cite{lindsay2011issues} with $p<\infty$. The case where the sub-likelihoods have diverging number of parameters  may be tackled by developing a more general penalty that discriminates parameters both within  and between sub-likelihoods. 

An open research issue is the choice of the tuning constant $\lambda$, which should ultimately depend on one's analysis objectives. Although we do not offer here a universal rule to select $\lambda$, standard  model-selection methods may be applied along the solution path, including information criteria and cross-validation procedures. Yet some care should be taken, since these strategies might not be intended for the large $p$-setting. Alternatively, one may consider a sequence of hypotheses tests for nested models   along the path, such as Wald, score, or likelihood-ratio  tests.

\section*{Acknowledgement} 

The authors acknowledge the financial support from Italian ministry MIUR -- Research project of national interest (PRIN) grant  2017TA7TYC.

%\section*{Supplementary material} 
%Supplementary material available at \emph{Biometrika} online includes technical proofs of the theorems in Section \ref{sec:properties}.

\section*{Appendix: technical proofs}

\subsection*{Proof of Theorem \ref{thm:consistency}}
\begin{proof}

Note that $\hat d(w)$ is strictly convex. If there exists a local minimizer of $\hat d(w)$ that is root-$(n/p)$ consistent,  by convexity  such a local minimizer must be $\hat w$. To show that there is a root-$(n/p)$ local minimizer of $\hat d_\lambda(w)$, we prove that for any $\epsilon >0$ there exists a large constant $C$ such that
\begin{equation}\label{eq_local}
\underset{n}{\lim \inf} \ P \left\{ \underset{\Vert v \Vert_2 = C}{\inf}  \hat d\left( w^\ast + \left(\dfrac{p}{n}\right)^{\frac{1}{2}} v \right) > \hat d\left( w^\ast \right) \right\} > 1-\epsilon.
\end{equation}
Define $\hat h= \text{diag}(\hat C)$ and consider the difference 
\begin{align} \nonumber
 & \hat d\left( w^\ast + \left(\dfrac{p}{n}\right)^{\frac{1}{2}} v \right) - 
  \hat d\left( w^\ast \right) \notag \\ \nonumber
= & \dfrac{1}{2} \left( w^\ast + \left(\dfrac{p}{n}\right)^{\frac{1}{2}} v\right)^\top \hat C \left( w^\ast + \left(\dfrac{p}{n}\right)^{\frac{1}{2}} v\right) - \left( w^\ast + \left(\dfrac{p}{n}\right)^{\frac{1}{2}} v\right)^\top \hat h - \dfrac{1}{2} w^{\ast \top} \hat C w^\ast + w^\ast \hat h  \notag \\ \nonumber
 & \ \ + \dfrac{\lambda}{n} \sum_{j=1}^p \dfrac{\left\vert w^\ast + \left(p/n\right)^{1/2}u_j\right\vert}{\tilde \theta_j} -  \dfrac{\lambda}{n} \sum_{j=1}^p \dfrac{| w^\ast_j|}{\hat \theta_j}\\ \nonumber
\geq & \dfrac{p}{2n}  v^\top \hat C v - \left(\dfrac{p}{n} \right)^{1/2}u^\top \left( \hat h -\hat C w^\ast \right) - \dfrac{\lambda}{n} \sum_{j=1}^{p^\ast} \dfrac{\left\vert w^\ast + \left(p/n\right)^{1/2}u_j\right\vert - \left\vert w^\ast \right\vert}{\tilde \theta^2_j}\\ \nonumber
\geq & \dfrac{p}{2n}  v^\top \hat C v - \left(\dfrac{p}{n} \right)^{1/2}u^\top \left( \hat h -\hat C w^\ast \right) - \left( \dfrac{p}{n}\right)^{1/2} \dfrac{\lambda}{n} \sum_{j=1}^p \dfrac{\left\vert v_j \right\vert}{\tilde \theta^2_j}\\
= & I_1 - I_2 - I_3\,. \label{eq:terms}
\end{align}
For $I_1$, we have
\begin{align} \label{eq:I1_bound}
I_1 = \dfrac{p}{2n}  v^\top \hat C v \geq \dfrac{p}{2n} \hat \rho_{\text{min}} \Vert v \Vert^2_2 =  O_p\left( \dfrac{p}{n} \right) k_1 \Vert v \Vert^2_2,
\end{align}
where $\hat \rho_{\text{min}}$  is the smallest eigenvalue of $\hat C$ and $\hat \rho_{\text{min}} = k_1 + o_p(p/n)$, $k_1 > 0$, by assumption. %\Ccom{??}. 
For $I_2$, applying the Cauchy-Schwartz inequality gives   
\begin{align}
I_2 & \leq  \left(\dfrac{p}{n}\right)^{1/2}  \left\Vert \hat h - \hat C w^\ast \right\Vert_2 \Vert v \Vert_2  \notag \\
%& \leq  \left(\dfrac{p}{n}\right)^{1/2} \left( \left\Vert \hat h -  h \right\Vert_2 + \left\Vert C w^\ast - \hat C w^\ast \right\Vert_2 \right) \Vert v \Vert_2 \notag \\
& \leq  \left(\dfrac{p}{n}\right)^{1/2} \left( \left\Vert \hat h -  h \right\Vert_2 + \left\Vert C w^\ast - \hat C w^\ast \right\Vert_2 \right) \Vert v \Vert_2 = O_p \left(\dfrac{p}{n}\right) \Vert v \Vert_2\,, \label{eq:I2_bound}
\end{align}
%\Ccom{last and second to last lines are the same}
where the last equality follows from $\Vert w^\ast \Vert \leq \Vert w^\ast \Vert_1 <  c$ for some constant $c<\infty$. %({\bf Need lemma here}) 
%and Assumption \Ccom{??}. 
For the last term $I_3$, we have 
\begin{align}
I_3 & \le \left( \dfrac{p}{n}\right)^{1/2} \dfrac{\lambda}{n} \left(\sum_{j\in \cA} |\tilde \theta_j|^{-2}\right)^{1/2} \left\Vert v \right\Vert_2 \notag \\
& \leq \left( \dfrac{p}{n}\right)^{1/2} \dfrac{\lambda}{n} \dfrac{\sqrt{p^\ast}}{\min_{j \in \cA}  |\tilde \theta_j|   } \left\Vert v \right\Vert_2 \notag \\
& \leq 
\left( \dfrac{p}{n}\right)^{1/2} \dfrac{\lambda}{n} \dfrac{\sqrt{p}}{\min_{j \in \cA}  |\tilde \theta_j|   } \left\Vert v \right\Vert_2 \label{eq:I3_bound1}
\end{align}
where the first inequality follows from the Cauchy-Shwartz inequality. Note that Condition A1 implies 
 \begin{align*}
 \min_{j \in \cA} |\theta_j^\ast| \leq  \max_{j \in \cA} |\tilde \theta - \theta_j^\ast| + \min_{j \in \cA} |\tilde \theta_j^\ast| =  o_p(1) + \min_{j \in \cA} |\tilde \theta_j^\ast|.
 \end{align*} 
Thus, if $\lambda n^{-1/2} \rightarrow 0$, from (\ref{eq:I3_bound1}) we have 
\begin{align} \label{eq:I3_bound}
I_3 & \leq 
\left( \dfrac{p}{n}\right) \dfrac{\lambda}{n^{1/2}} \dfrac{1}{\min_{j \in \cA}  | \theta_j | + o_p(1) } \left\Vert v \right\Vert_2 = O_p\left( \dfrac{p}{n} \right) \left\Vert v \right\Vert_2.
\end{align}
Given (\ref{eq:I1_bound}), (\ref{eq:I2_bound}) and (\ref{eq:I3_bound}), the positive term $I_1$ dominates  $I_2$ and $I_3$ in (\ref{eq:terms})  when $\Vert v\Vert_2$ is  allowed to be large. This shows (\ref{eq_local}) and therefore completes the proof.
\end{proof}
\bibliographystyle{abbrvnat} 

\subsection*{Proof of Theorem \ref{thm:sparse_solution}}

We show that with probability tending to 1 $\hat w = (\hat w_{\cA}, 0^\top)^\top$, is the minimizer of $\hat d_{\lambda}(w)$. From the Karush-Kuhn-Tucker condition given in (\ref{kkt}), a necessary and sufficient condition is 
\begin{equation*}
P \left( \forall j \in \cA^c, \left\Vert 
\dfrac{1}{n}\sum_{i=1}^n  \tilde u^{(i)}_{j}(\tilde u^{(i)}_{j} - \hat w_{\cA}^\top \tilde u_{\cA}^{(i)}) 
\right\Vert_2 \le 
\dfrac{\lambda }{n \tilde{\theta}_j^{2}} 
\right) \rightarrow 1. 
\end{equation*}
Note that 
\begin{align*}
 \left\Vert \dfrac{1}{n}\sum_{i=1}^n   \tilde u^{(i)}_{j}(\tilde u^{(i)}_{j} - \hat w_{\cA}^\top \tilde u_{\cA}^{(i)}) 
\right\Vert_2 &  \le   \left( \dfrac{1}{n}\sum_{i=1}^n   (\tilde u^{(i)}_{j})^2 \right)^{1/2}+ \left\Vert \hat w_{\cA}^\top \dfrac{1}{n}\sum_{i=1}^n   \tilde u^{(i)}_{j}\tilde u_{\cA}^{(i)} 
\right\Vert_2  \\
& \leq C^{1/2}_{jj} + o_p(1) + p^\ast \left\vert \max_{k \in \cA} \dfrac{1}{n}\sum_{i=1}^n   \tilde u^{(i)}_{j}\tilde u_{k}^{(i)} 
\right\vert   \left\Vert \hat w_{\cA}\right\Vert_2
\end{align*}
We have that $\left\Vert \hat w_{\cA} \right\Vert_2 \leq \left\Vert \hat w_{\cA} \right\Vert_1 < c_1$ for some $c_1 \leq \infty$. Thus, by Condition A4, we have
\begin{align*}
 \left\Vert \dfrac{1}{n}\sum_{i=1}^n   \tilde u^{(i)}_{j}(\tilde u^{(i)}_{j} - \hat w_{\cA}^\top \tilde u_{\cA}^{(i)}) 
\right\Vert_2 
& \leq C^{1/2}_{jj} + O_p(1).
\end{align*}
Since $n \tilde{\theta}_j^{2}$ converges in probability to 1, $\lambda/n \tilde{\theta}_j^{2}$ diverges for any $j \in \cA^c$. Thus, the proof is complete.

\subsection*{Proof of Theorem \ref{thm:oracle}}

Theorem \ref{thm:sparse_solution} states that, with probability tending to 1, the minimizer of Criterion (\ref{eq:criterion_empirical}) is equal to $(\hat w^{\ast \top}_\cA, 0^\top)^\top$; this that the estimator $\hat \theta_j$ is exactly equal to zero for $j \in \cA^{c}$ with probability going to 1. Consequently, it remains to show that 
\begin{equation}\label{eq:modsel_consistency}
P\left( \min_{j \in \cA} |\hat \theta_j| > 0 \right) \rightarrow 1.
\end{equation}
Since the preliminary estimator $\tilde \theta_j$ is root-$n$ consistent, also the implied one-step estimator $\tilde \theta_j$ is root-$n$ consistent with $\max_{j \in \cA} |\hat  \theta_j - \theta_j^\ast| = o_p(1)$. Therefore,
\begin{align*}
 \min_{j \in \cA} |\hat \theta_j| \geq  \min_{j \in \cA} |\theta_j^\ast| - \max_{j \in \cA} |\hat  \theta_j - \theta_j^\ast| = \min_{j \in \cA} |\theta_j^\ast| + o_p(1)
 \end{align*} 
 and $\min_{j \in \cA} |\theta^\ast_j|>0$ imply (\ref{eq:modsel_consistency}), so the proof   is complete.

\bibliography{biblio}

\end{document}